\documentclass[aps,prb,twocolumn,groupedaddress]{revtex4}
\usepackage[pdftex]{graphicx}

\begin{document}
\title{Gigahertz quantized charge pumping in graphene quantum dots}

\author{M. R. Connolly$^{1,2}$, K. L. Chiu$^2$, S. P. Giblin$^1$, M. Kataoka$^1$, J. D. Fletcher$^1$, C. Chua$^2$, J. P. Griffiths$^2$, G. A. C. Jones$^2$, V. I. Fal'ko$^3$, C. G. Smith$^2$ \& T. J. B. M. Janssen$^1$}
\affiliation{$^1$National Physical Laboratory, Hampton Road, Teddington TW11 0LW, UK}
\affiliation{$^2$Cavendish Laboratory, Department of Physics, University of Cambridge, Cambridge, CB3 0HE, UK}
\affiliation{$^3$Department of Physics, Lancaster University, Lancaster, LA1 4YB, UK}

\maketitle

\textbf{Single electron pumps are set to revolutionize electrical metrology by enabling the ampere to be re-defined in terms of the elementary charge of an electron \cite{Zimmerman2003}. Pumps based on lithographically-fixed tunnel barriers in mesoscopic metallic systems \cite{Keller1996, Keller1999} and normal/superconducting hybrid turnstiles \cite{Pekola2008,averin2008} can reach very small error rates, but only at MHz pumping speeds corresponding to small currents of the order 1~pA. Tunable barrier pumps in semiconductor structures have been operated at GHz frequencies \cite{fujiwara2008nanoampere,Giblin2012}, but the theoretical treatment of the error rate is more complex and only approximate predictions are available \cite{kashcheyevs2010universal}. Here, we present a monolithic, fixed barrier single electron pump made entirely from graphene. We demonstrate pump operation at frequencies up to 1.4 GHz, and predict the error rate to be as low as 0.01 parts per million at 90 MHz. Combined with the record-high accuracy of the quantum Hall effect \cite{Tzalenchuk2010} and proximity induced Josephson junctions \cite{Jeong2011}, accurate quantized current generation brings an all-graphene closure of the quantum metrological triangle within reach \cite{Piquemal2000}. Envisaged applications for graphene charge pumps outside quantum metrology include single photon generation via electron-hole recombination in electrostatically doped bilayer graphene reservoirs \cite{Mueller2009a}, and for read-out of spin-based graphene qubits in quantum information processing \cite{Trauzettel2007}.}

\begin{figure}[!t]
\includegraphics{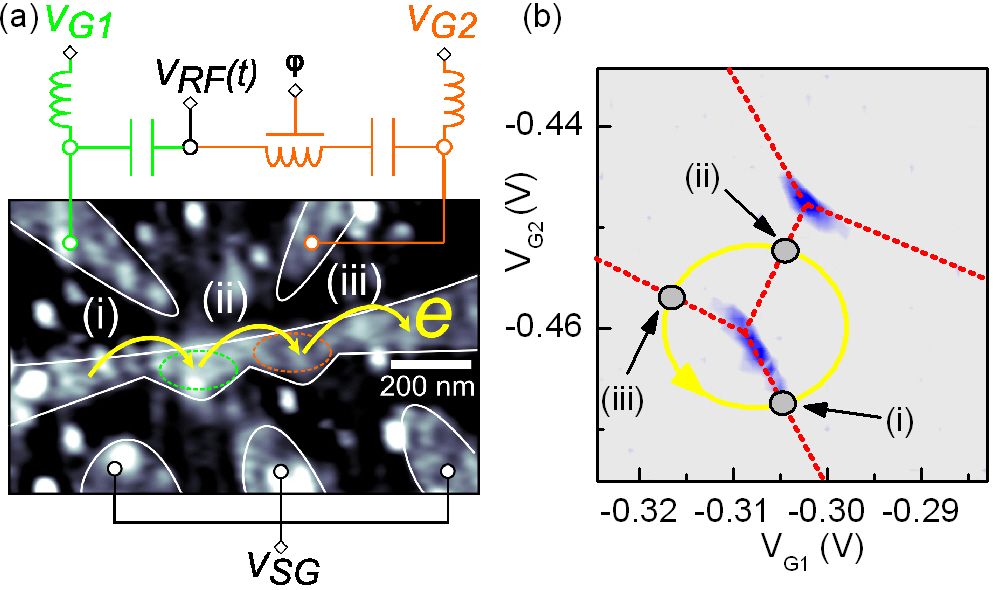}
\caption{\textbf{Adiabatic pumping mechanism in a graphene double quantum dot.} \textbf{(a)} Atomic force microgaph of the device and experimental setup used to measure and generate the pumped current. An oscillating voltage $V_{RF}(t)$ is added to the DC voltages $V_{G1}$ and $V_{G2}$, which control the total number of electrons on each quantum dot. The phase delay $\phi$ added between the modulation voltage applied to $V_{G1}$ and $V_{G2}$ controls the shape of the pump loop in the space of the two gate voltages. \textbf{(b)} DC current through the device as a function of DC voltages applied to the gates. Red dashed lines indicate the edges of the honeycomb stability diagram (See Supplementary Section S\ref{Locating triple points suitable for optimised pumping}). A quantized current is pumped when the gate voltage modulation produces a trajectory (yellow) that encircles a triple point, passing through the sequence of transitions (i)$\rightarrow$(ii)$\rightarrow$(iii).}    
\label{Fig:Fig1}
\end{figure}

The graphene double quantum dot pump used in this work consists of two lithographically defined graphene islands which are coupled to each other and to source and drain contacts by narrow constrictions [(Fig. \ref{Fig:Fig1}(a)]. The strong Coulomb interaction between electrons blocks the continuous flow of current through the device until nearby plunger gates lower the energy cost of tunneling through both quantum dots. To pump electrons from source to drain we rapidly modulate the voltage on the gates such that only a single electron can be transfered through the structure per modulation cycle \cite{Pothier1992}. A single cycle comprises the three stages which are illustrated in Fig. \ref{Fig:Fig1}: (i) the electrostatic potential on the first quantum dot is lowered as the voltage $V_{G1}$ on plunger gate 1 increases, pulling an electron in from the source lead; (ii) the potential on the second quantum dot is lowered as the voltage $V_{G2}$ on plunger gate 2 increases, shifting the electron to the second dot; (iii) the electron is pushed out to the drain contact and the system returns to its original charge configuration ready to pump the next electron. The frequency $f$ of the oscillating voltage $V_{RF}$ applied to the gates determines the rate at which electrons are transferred, and thus the size of the pumped current, $I=ef$.

\begin{figure}[!b]
\includegraphics[width=89mm]{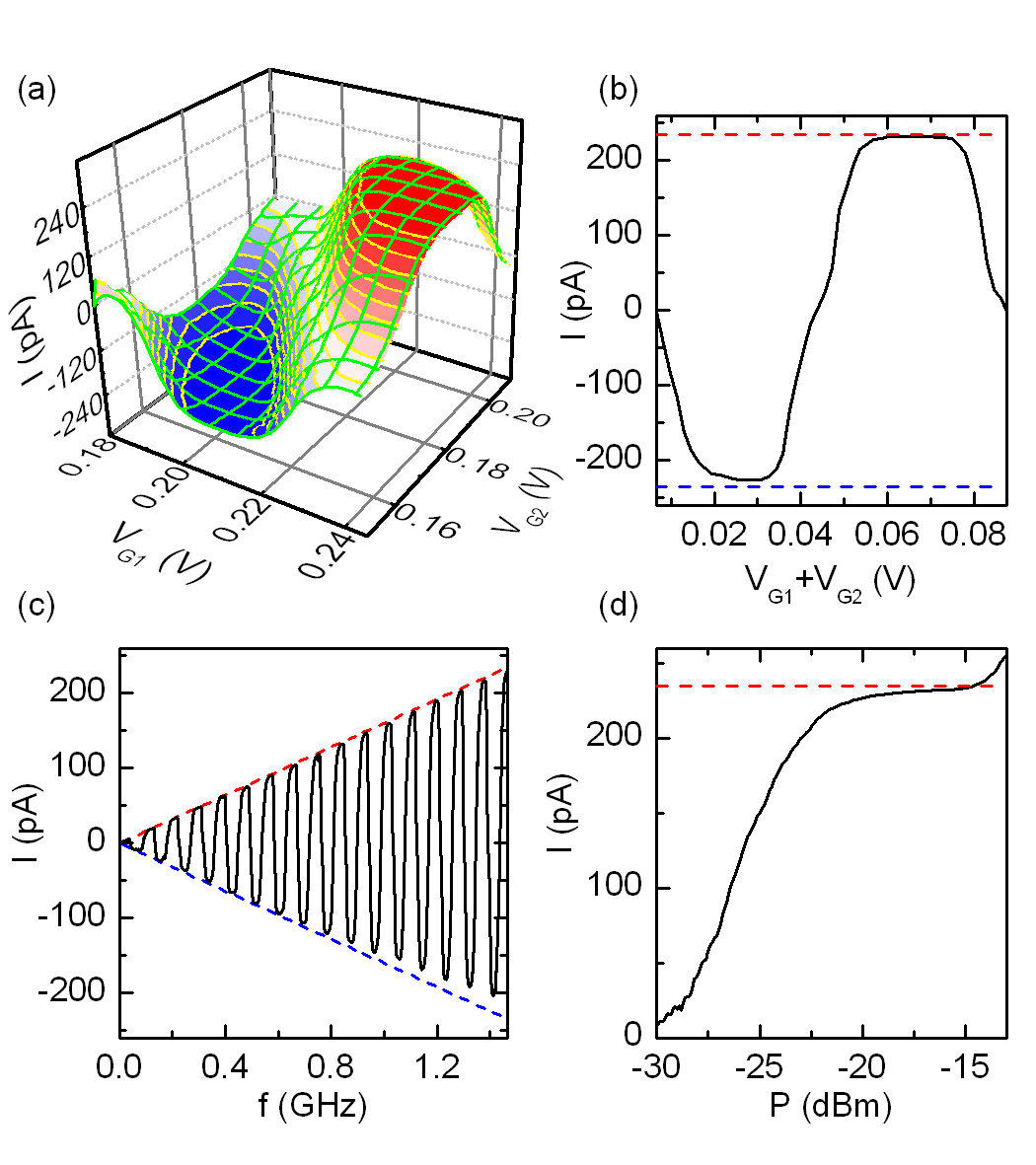}
\caption{\textbf{Gigahertz quantized charge pumping.} (T=300 mK) \textbf{(a)} Pumped current generated by the double quantum dot as a function of $V_{G1}$ and $V_{G2}$ at $f=$1.465 GHz ($ef\approx$235 pA) and $P$=-15 dBm. \textbf{(b)} Plot showing the pumped current along a line passing through both the triple points. \textbf{(c)} Pumped current as a function of frequency ($P$=-15 dBm) and \textbf{(d)} power ($f=$1.465 GHz). Red and blue dashed lines show the quantized values $I=\pm ef$ of the pumped current.}    
\label{Fig:Fig2}
\end{figure}

Figure. \ref{Fig:Fig2}(a) shows the pumped current generated by applying a modulating voltage at $f\approx$ 1.5 GHz while sweeping the voltages $V_{G1}$ and $V_{G2}$ in the vicinity of a pair of triple points (Supplementary Fig. S\ref{Fig:FigS2}). We observe extended plateaus in the pumped current at equal and opposite values around each triple point [Fig. \ref{Fig:Fig2}(b)], demonstrating that the charge pump delivers single electrons from source to drain per cycle. The opposite sign of the pumped current when encircling different triple points in a pair is expected, as the equivalent sequence of configurations pumps an electron in the opposite direction, or holes in the same direction \cite{VanDerWiel2002}. Unambiguous confirmation of quantized charge pumping is shown in Fig. \ref{Fig:Fig1}(c), which plots the pumped current as a function of $f$ with the DC voltages fixed on the right hand triple point. Aside from the oscillatory behaviour, which is introduced as the pump loop periodically changes direction around the triple point (Supplementary Section S\ref{Oscillations in sign of pumped current with frequency}), we observe remarkable adherence to the quantized value $I=\pm ef$ over a range of frequencies up to a few GHz. At fixed pump frequency and gate voltage we also observe a plateau in the pumped current as a function of the RF power controlling the size of the pump loop [Fig. \ref{Fig:Fig2}(d)]. The independence of pumping efficiency with power confirms that our pump loop can satisfy the purely topological requirement of encircling a triple point. It is likely that the departure from $ef$ for negative currents derives from leakage currents due to intersecting triple points in the vicinity of the left triple point [See Supplementary Section S\ref{Locating triple points suitable for optimised pumping}].

\begin{figure}[!b]
\includegraphics[width=89mm]{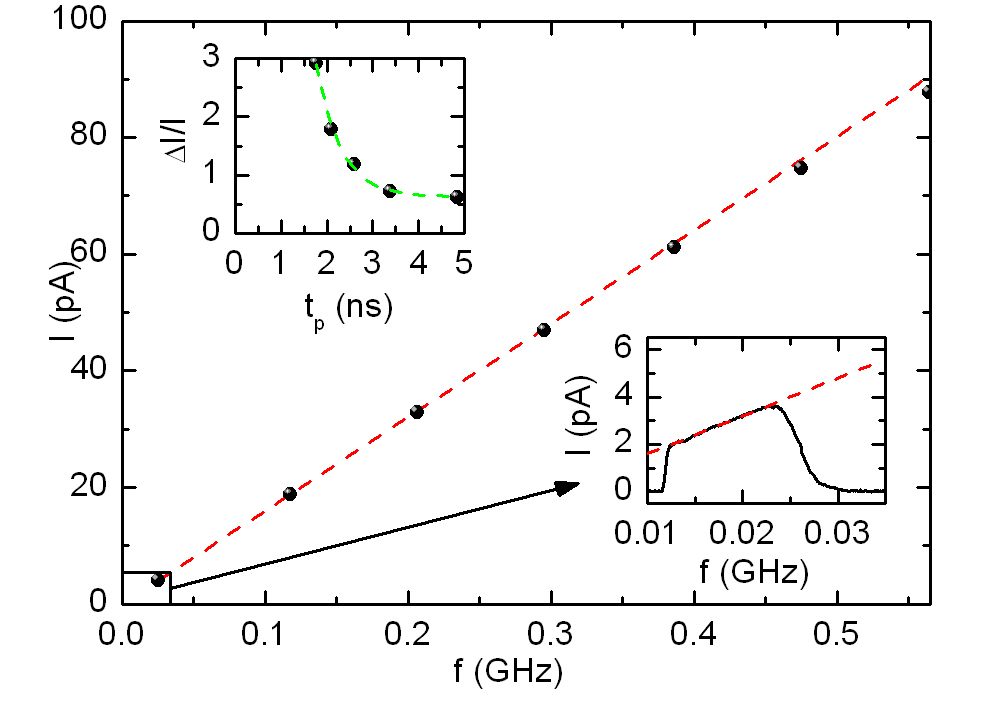}
\caption{\textbf{Frequency dependence of the pumped current.} Current as a function of drive frequency (1.2 K); Dashed line (red) is the linear relation expected from pumping one electron per cycle. The lower right inset shows the low frequency pumped current and the upper left inset shows the percentage error in the pumped current as a function of the pump cycle period. See Supplementary Section S\ref{Oscillations in sign of pumped current with frequency} for explanation of the frequencies where the current is not quantized. A systematic offset of $~$40 fA due to the uncalibrated electrometer has been subtracted from the raw data.}    
\label{Fig:Fig3}
\end{figure}

\begin{figure*}[!t]
\includegraphics{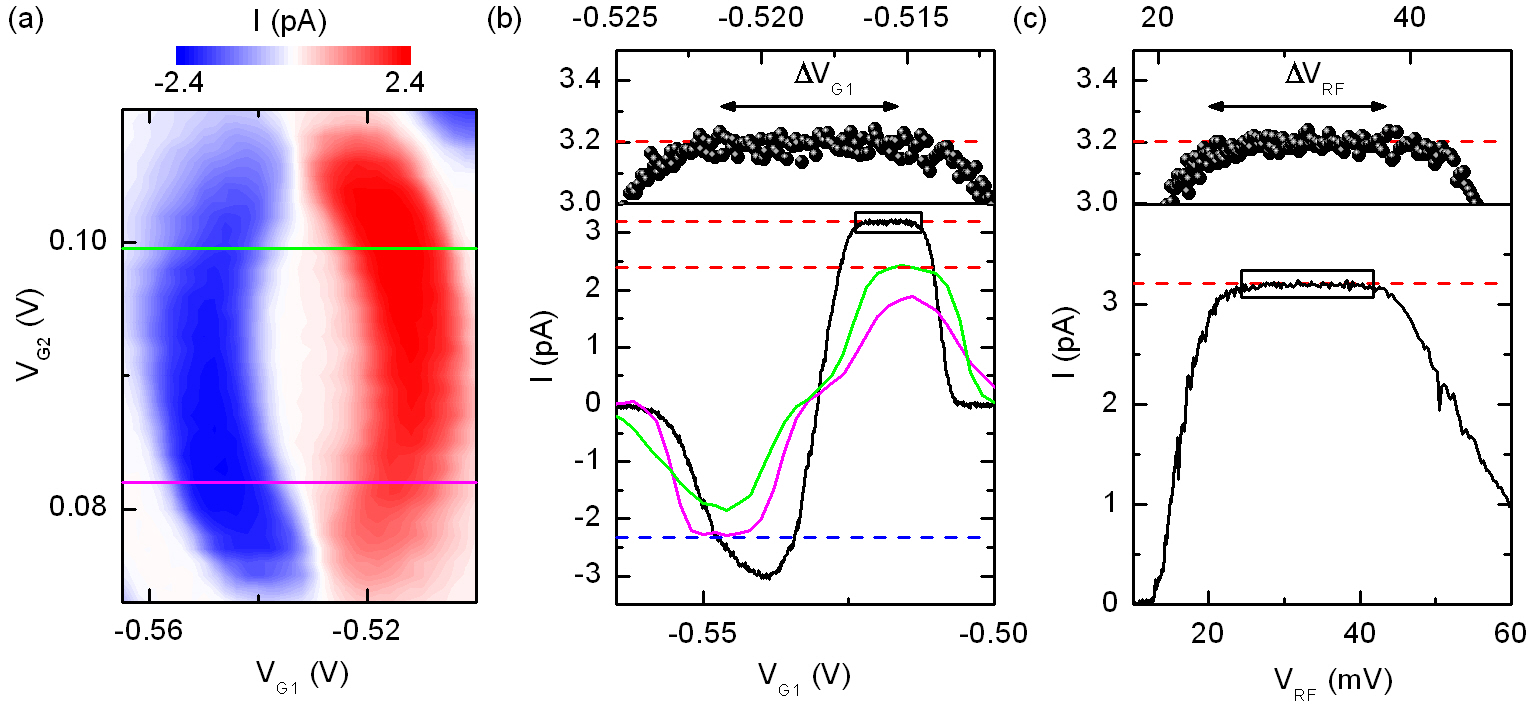}
\caption{\textbf{Quantization accuracy at low frequency}. \textbf{(a)} Map of the pumped current as a function of the DC voltages $V_{G1}$ and $V_{G2}$ while modulating $V_{RF}$ at $f=$15 MHz. \textbf{(b)} Line profiles of the pumped current as a function of $V_{G1}$ at $f=$15 MHz (pink, green) and $f=$20 MHz (black). The red/blue dashed lines are the values expected from pumping one electron/hole per cycle. \textbf{(c)} Pumped current as a function of the pump loop size with the gate voltages fixed next to the right-hand triple point.}    
\label{Fig:Fig4}
\end{figure*}

Owing to the demand for high currents in the majority of charge pump applications, a key figure of merit is the quantization accuracy at high frequency \cite{Blumenthal2007}. The speed of tunneling through potential barriers illustrated in Fig. \ref{Fig:Fig1} is limited by the $RC$ time constant of the barriers. To ensure that quantum fluctuations do not destroy charge quantization on the islands, $R\gg R_{\text{K}}$. Conventional metal-oxide junctions formed by electron beam lithography have capacitances of order $1$~fF and therefore $RC\sim 0.1$~ns. In pumps formed from chains of these metallic junctions, the error rate due to missed tunnel events can be as low as 0.01 parts per million for $f<10$~MHz, but is expected to degrade exponentially with increasing frequency, approaching $1$ for $f=1$~GHz \cite{Martinis1994,Keller1996}. On the other hand, tunnel barriers formed in silicon \cite{zimmerman2006electrostatically,yamahata2008control} and graphene \cite{Stampfer2008a} have capacitances at least an order of magnitude less than metal-oxide barriers with the same tunnel resistance, leading to faster $RC$ time constants, although high speed quantized pumping in a semiconductor fixed-barrier pump has not previously been demonstrated. In our graphene device we estimate $R\sim 100$~k$\Omega$ from the differential resistance at high bias voltage, and $C\sim 40$~aF from the Coulomb charging energy (Supplementary Section S\ref{Location of quantum dots}). 

In Fig. \ref{Fig:Fig3} we show the pumped current as a function of frequency for a different pair of triple points to the data in Fig. \ref{Fig:Fig2}. At low frequency (lower right inset), the current is equal to $ef$, within the $\sim 0.2 \%$ calibration accuracy of our ammeter, but at high frequency the current falls below $ef$, a behaviour we attribute to an increase in the number of failed pumping cycles. In the upper left inset we plot the error $\Delta I/I = (ef-I)/I$ as a function of cycle time $t_p=1/f$, for the higher frequencies where the error is substantially larger than the measurement uncertainty. A fit of this data to the expected exponential dependence $\Delta I/I \propto exp[-t_p /\tau]$ yields $\tau\simeq 0.6$~ns. Similar fits to error measurements on 5-junction \cite{Martinis1994} and 7-junction \cite{Keller1996} metallic pumps yielded $\tau\simeq 6$~ns. This scaling of the observed error rates is expected from the roughly $10$ times scaling of the junction capacitances, and extrapolating to lower frequencies, we predict an error rate due to missed tunnel events of less than 0.01 ppm at $f=90$~MHz.  We note that the error rate due to missed tunnel events depends only weakly on the number of junctions in the pump, in contrast to co-tunneling errors which are strongly suppressed by adding extra junctions \cite{Jensen1992}.

In Fig. \ref{Fig:Fig4} we examine in more detail the low-frequency pumping and its robustness against variations in control parameters. The pump map around one pair of triple points at $f$=15 MHz is shown in Fig. \ref{Fig:Fig4}(a). The phase difference between $V_{G1}$ and $V_{G2}$ was optimized so as to avoid any DC current contribution (see Supplementary Section S\ref{Locating triple points suitable for optimised pumping} for more details of this procedure.) Figure \ref{Fig:Fig4}(b) shows comparisons between the line sections $I(V_{G1})$ passing through each triple point and and the values $I$=$\pm ef$ expected if one electron is transfered per cycle. Flat plateaus in $I(V_{G1})$ persists with an average value of 3.189 pA and the extent of the plateau is $\Delta V_{G1}$$\approx$6 mV, which is close to the average TP spacing of $\Delta V_{TP}$$\approx$ 5.5 mV [see Supplementary Fig. S\ref{Fig:FigS1}]. Fig. \ref{Fig:Fig4}(c) demonstrates how robust this quantization is against variations in the size of the pump loop. At low $V_{RF}$ the pump loop neither encloses or passes through a triple point, nor intersects the cotunneling and polarization lines spanning between them, so the time-averaged current is close to zero. As the pump loop expands it first passes through a TP and the current begins to increase. Once beyond, the pump loop encircles the triple point and a current plateau develops and extends $\Delta P$$\approx$3 dBm, or $\Delta V_{RF}$$\approx$12 mV. At high power the pump loop encloses both triple points and the current consequently decreases. These plateaus suggest the pump is robust against low-level charge fluctuations occuring over several hours of normal pump operation, although ocassional instabilities did occur in certain ranges of gate voltage (see Supplementary Fig. S\ref{Fig:FigS3}). Such instabilities in the charge configuration of the dots due to charge fluctuations in the SiO$_{2}$ can be minimized by using boron nitride \cite{Mayorov2011} substrates which are free from dangling bonds and charge traps \cite{Dean2010}.

We ascribe the high performance of our adiabatic pumps to a number of factors deriving from graphene's unique two-dimensional physical and electronic structure. Firstly, the presence of strong edge and potential disorder in lithographically defined graphene nanostructures leads to the formation of multiple quantum dots in the constrictions acting as tunnel barriers between the dots (Supplementary Section S\ref{Role of disorder}). Rather than this impeding pumping, the resulting capacitance and resistance of the random tunnel junctions between the quantum dots promotes a high intrinisic tunneling time while simultaneously suppressing cotunneling events due to the large overall dissipation \cite{Keller1996, Zorin2000}. Secondly, the large interdot capacitance and linear electronic dispersion in graphene leads to a large and occupancy dependent single-particle energy spacing near the Dirac point, $\Delta (N)=\hbar v_F/(d \sqrt{N})$ ($N<5$) \cite{Molitor2009}, which suppresses photon assisted interdot transitions \cite{Gasser2009} and protects adiabaticity of the pump when operated at high frequency. This permits the interdot capacitance to be optimized through increased dot size, thereby improving the tolerances for parallelization, without compromising on accuracy. The large separation between triple points in strongly capacitively coupled dots also suppresses leakage currents while crossing the polarization line of the pumping cycle [(ii)$\rightarrow$(iii) in Fig. \ref{Fig:Fig1}(b)]. In future, the precise origin of error mechanisms at high frequency and the role of cotunneling can be tested by equipping our devices with charge detectors \cite{Keller1996,Guttinger2011a}. 

Our predicted performance of 0.01 ppm at $\approx$100 MHz implies ten pumps operated in parallel would deliver 100 pA with metrological accuracy. The monolithic processing offered by large-area graphene grown by decomposition of silicon carbide \cite{Virojanadara2008, Emtsev2009, Tzalenchuk2010} or chemical vapour deposition \cite{Reina2009} is ideally suited for such large-scale parallelization and integration with other graphene nanoelectronic devices. In particular, interfacing charge pumps with edge states in the quantum Hall regime would lead to a high accuracy voltage standard operable at high temperature and lower magnetic fields \cite{Hohls2011,Tzalenchuk2010, Janssen2011}. Since induced superconductivity and Josephson effects \cite{Jeong2011} persist in graphene at magnetic fields compatible with the quantum Hall effect \cite{Rickhaus2012}, a realization of the quantum metrological triangle in a single graphene device is also now within sight. 

The availability of graphene charge pumps will improve the prospects for graphene in such areas as single electron logic and quantum light generation. Pumping single electrons into gate defined $p$- and $n$-type regions of gapped bilayer graphene \cite{Zhang2009} would generate a stream of tuneable single photons via electron-hole recombination \cite{Mueller2009a}. Because of the close analogy between Dirac Fermions in graphene and photons, the single-electron charge pumps presented here could play a role in electron optics analogous to a single photon source in quantum optics. Combined with the ballisticity of electrons in suspended graphene or graphene on boron nitride \cite{Mayorov2011}, this would enable testing optics elements, such as beam splitters and Veselago lenses \cite{Cheianov2007}, in the single-electron regime. Our pumps can be tested against quantum pumping concepts which exploit graphene's unique bandstructure to achieve more exotic single charge and spin manipulation devices \cite{Prada2009,Liu2011}, and also pave the way towards the high frequency manipulation of charge required for developing graphene spin qubits \cite{Trauzettel2007}.

This work was financially supported by the European GRAND project (ICT/FET, Contract No. 215752) and a EPSRC/NPL Joint Postdoctoral Partnership. 

\section{Supplementary Information}

\subsection{Locating triple points suitable for optimised pumping}
\label{Locating triple points suitable for optimised pumping}
\begin{figure*}[!h]
\includegraphics{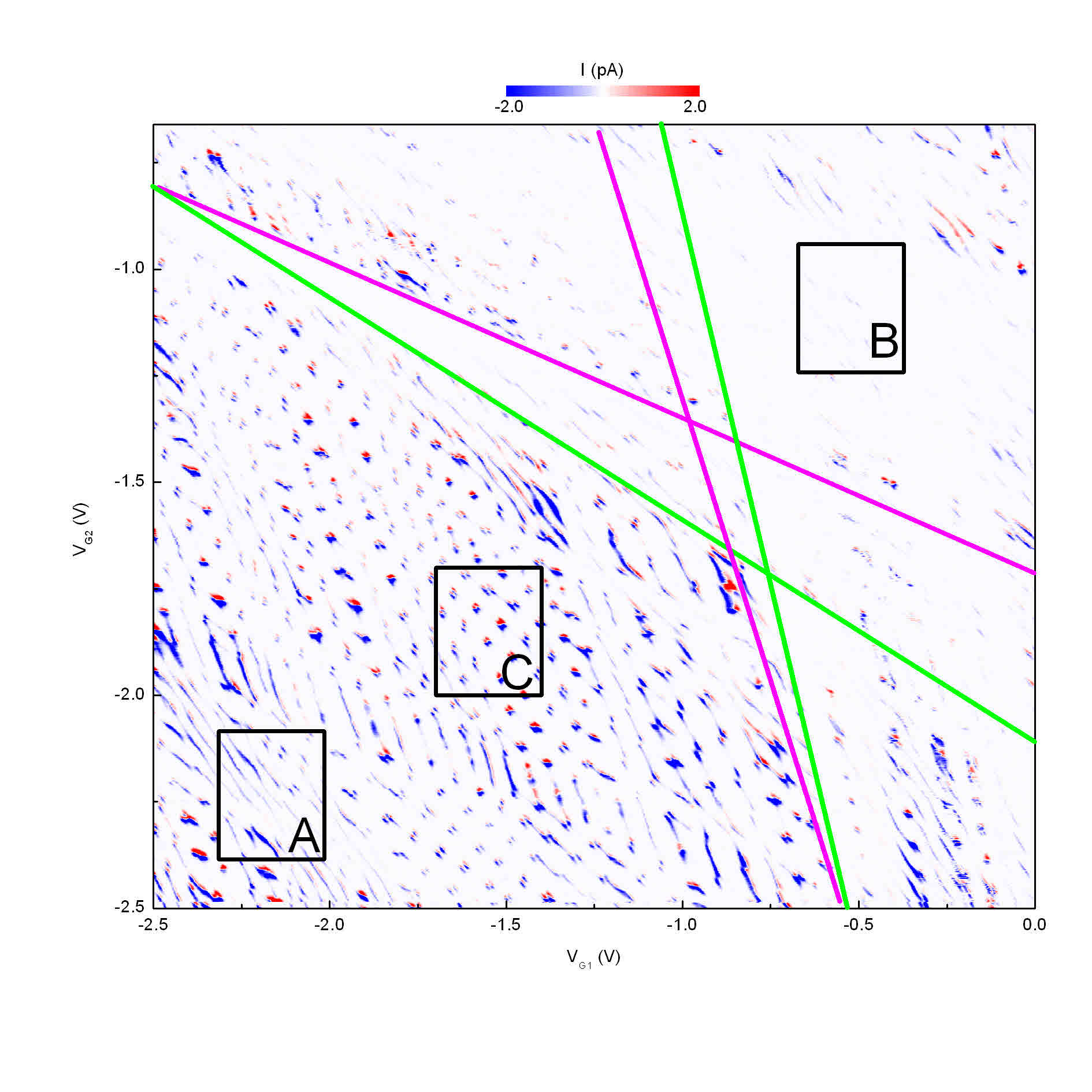}
\caption{\textbf{Zero-bias stability diagram}. ($T$=1.2 K) Outlines shows the regions where (A) pronounced single dot behaviour is observed due to the enhanced transparency of the interdot barrier, (B) the barriers are opaque, and (C) double dot behaviour dominates. The triple points shown in Fig \ref{Fig:Fig1}(b) are captured in the vicinity of region (B). Pink lines run parallel to the resonances in the barriers which give rise to the different tunnel coupling strengths, and green lines indicate the slope of the single-particle states.}    
\label{Fig:FigS2}
\end{figure*}

\begin{figure*}[!h]
\includegraphics{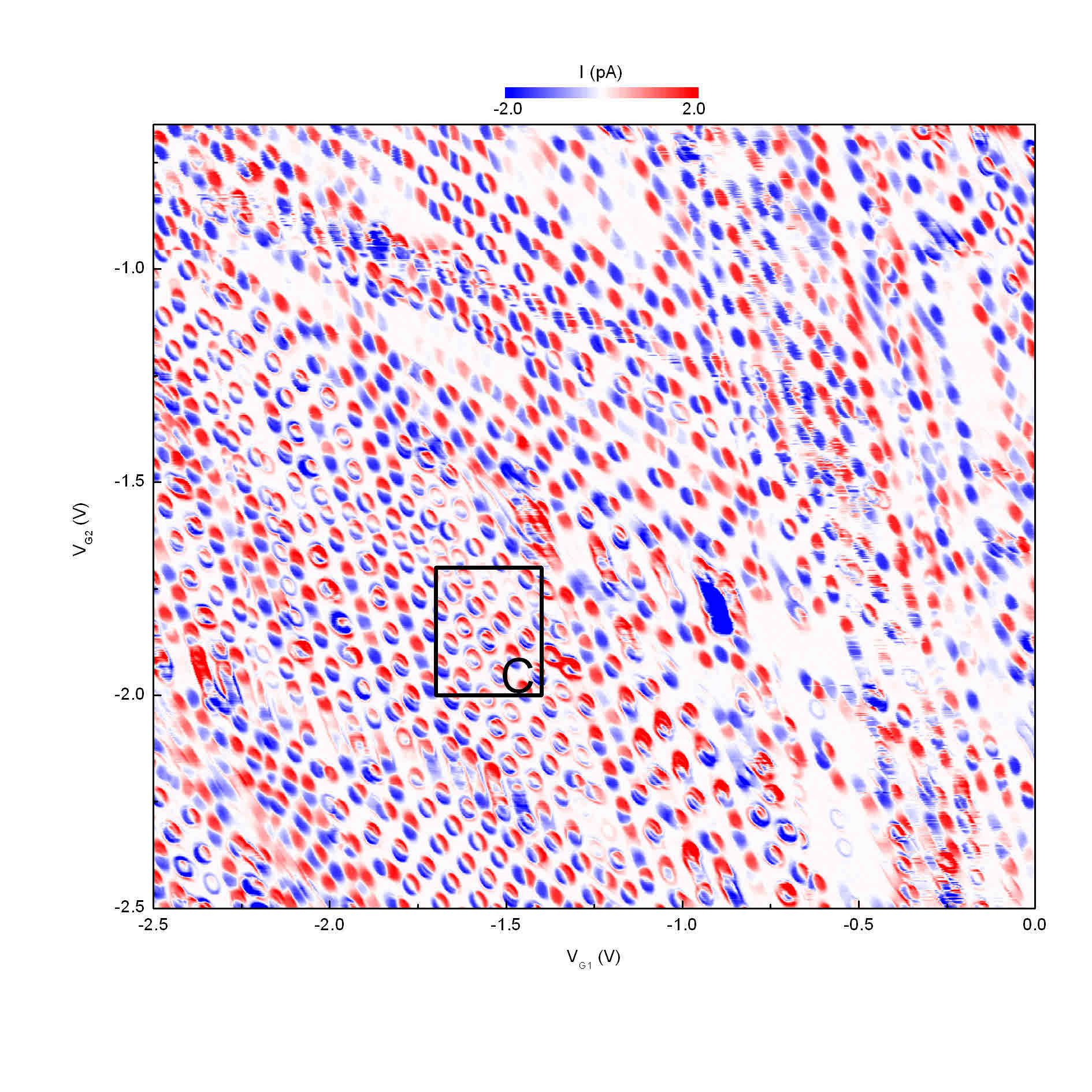}
\caption{\textbf{Quantized current stability diagram}. ($f$=12 MHz, $T$=1.2 K) Outline shows the region in Fig. S\ref{Fig:FigS1}(d).}    
\label{Fig:FigS3}
\end{figure*}

As discussed in the main text, the excursion ($N_{1}$, $N_2$)$\rightarrow$($N_{1}$+1, $N_2$)$\rightarrow$($N_{1}$+1, $N_2$)$\rightarrow$($N_{1}$, $N_2$+1) in gate-voltage space is all that is topologically required to pump a single electron through the double quantum dot (DQD) per cycle, where $N_1$ and $N_2$ are the number of electrons on quantum dot 1 (QD1) and quantum dot 2 (QD2), respectively. However, for accurate quantized pumping it is also necessary to avoid regions of the stability diagram where there is a background (non-pumped) contribution to the total current, i.e., upon crossing the edges of the honeycomb during stages (1) and (3), and between the triple points (TPs) during stage (2) [see Fig. \label{Fig:FigS1}(c)]. To find a suitable range of gate voltages where it is possible to satisfy these conditions by tuning the shape of the pump loop, prior to pumping we fix the side gate at the charge neutrality voltage \cite{Liu2010,Molitor2010a,Moriyama2009} and measure the DC current as a function of the DC voltages $V_{G1}$ and $V_{G2}$. Fig. S\ref{Fig:FigS2} illustrates the variability of this DC contribution at $T$=1.2 K, over a $\approx$2 V range of the gate voltages, corresponding to $\approx$30 single-particle states. We find large variations in both the size of the high current regions close to the vertices of the honeycomb cells and the low current ridges spanning their edges. In particular, at large negative voltages (Fig. S\ref{Fig:FigS2}, region A) the anti-crossings of the honeycomb structure are less pronounced, reflecting stronger inter-dot tunnel coupling, while at lower voltages (region B) the edges of the honeycomb are not visible, indicating cotunneling is strongly suppressed. These features of the stability diagram have been observed frequently in graphene DQDs \cite{Molitor2010a,Liu2010,Wang2012} and are attributed to the changing transparency of resonant states in the constrictions, which form in the narrow barriers due to potential and edge disorder and quantum confinement \cite{Han2007,Molitor2009,Todd2009,Molitor2010a,Stampfer2009,Liu2010,Gallagher2010,Terres2011}. Their presence is betrayed by the fact that weak and strong coupling regimes lie along bands (pink lines) roughly parallel to the DQD resonances (green lines). Such bands are naturally interpreted as broad resonances of smaller localized states \cite{Stampfer2008}, which have similar lever arms due to their symmetrical position and proximity to the gates. Their impact on operating the DQD as a charge pump is to exclude certain regions of configuration space where, for instance, single-dot behaviour is observed and the time-averaged DC current dominates. The pumped current over the same 2 V of gate voltages is shown in Fig. S\ref{Fig:FigS3}. Although regions where the barrier states are more opaque consist of absent and irregular TPs, the pumped current is well quantized owing to the suppressed cotunneling (absence of current along edges of the honeycomb), and is achieved without too much optimization of the pump loop. A TP pair from this region was used in the high frequency pumping measurement shown in Fig. \ref{Fig:Fig2}(c).

In the regions where the honeycomb is better developed the precise trajectory taken in configuration space, which is determined by the amplitude $V_{RF}$ of the drive and the phase shift $\phi$ between the two gates, has to be finely tuned. We select region C in Fig. S\ref{Fig:FigS3} where the honeycomb lattice is uniform to illustrate the pumping mechanism in Fig. S\ref{Fig:FigS1}. With the AC drive on the gates ($f$=12 MHz, $P$=-25 dBm), we observe overlapping ellipses with crescent-shaped areas of positive and negative current that develop around the left and right TPs of each pair in the honeycomb [Fig. S\ref{Fig:FigS1}(d)]. To understand the form and structure of these features, Fig. S\ref{Fig:FigS1}(e) shows a direct comparison of the locations in gate space around a typical pair of TPs. The flat regions ($P_+$ and $P_-$) within the crescents coincide with where the pump loop avoids the DC contributions, whereas the surrounding regions exhibit a current $|I|$$<$$ef$ owing to the additional time-averaged current.  The central region ($P_0$) where $I$$\approx$0 corresponds to encircling both TPs, leading to repeatedly increasing and decreasing the occupancy of each QD without any net transfer of electrons from source to drain. Changing the relative phase $\phi$ between the gates alters the trajectories around the TPs and consequently the DC contribution to the pumped current.

\begin{figure*}[!h]
\includegraphics{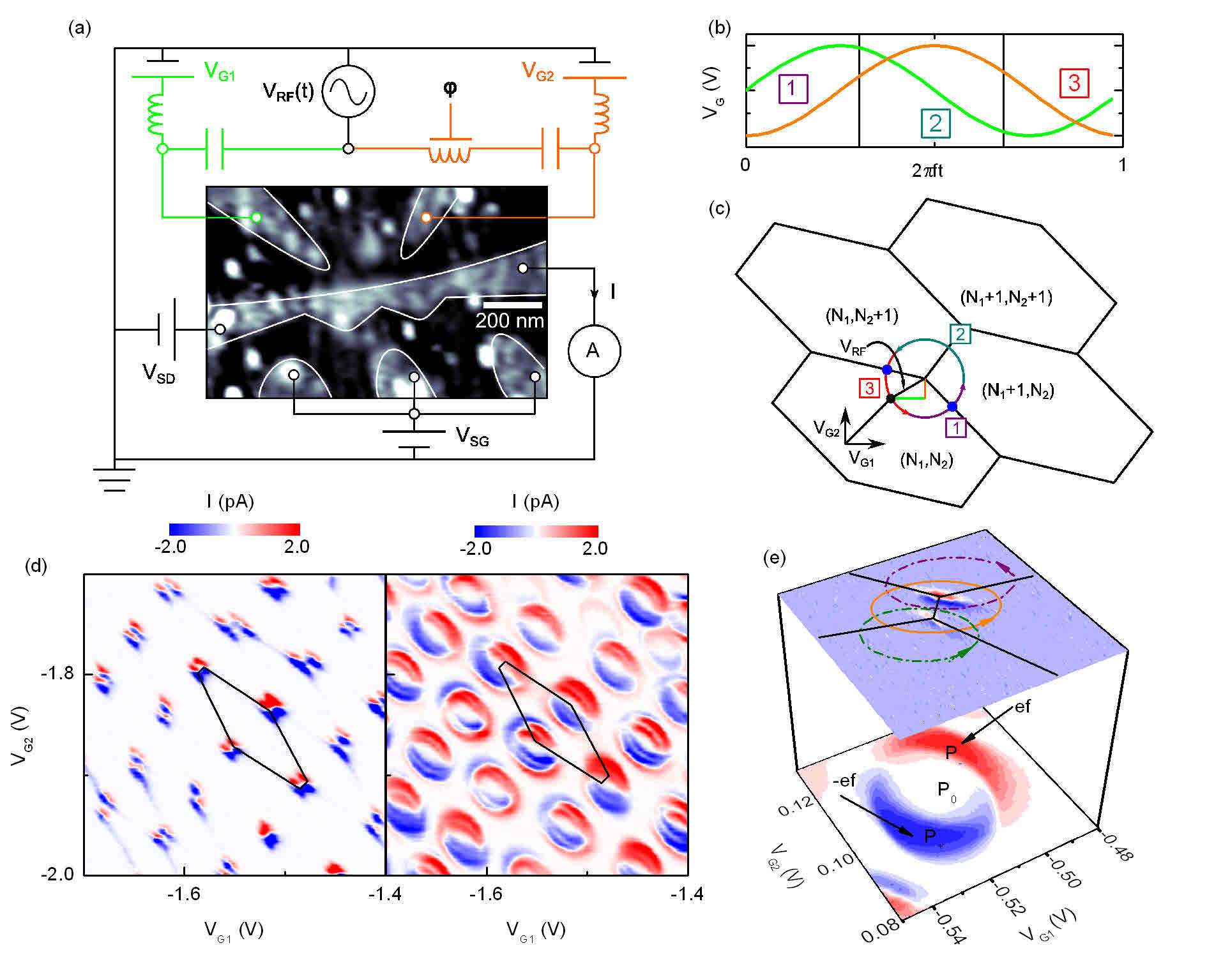}
\caption{\textbf{Experimental setup}. \textbf{(a)} Atomic force microgaph of the device and experimental setup used to measure and generate the pumped current. An RF signal is added to the DC voltages $V_{G1}$ and $V_{G2}$ used to control the potential on QD$_1$ and QD$_2$. A phase delay $\phi$ is added to the modulating signal applied to plunger gate 2 and controls the trajectory followed in gate space. \textbf{(b)} Plot showing the amplitude of the two plunger gate voltages and the three stages where charge rearrangements occur. \textbf{(c)} Double quantum dot stability diagram showing the honeycomb cells where charge configurations are stable. \textbf{(d)} Current through the DQD in region C of Figs. S\ref{Fig:Fig2},S\ref{Fig:Fig3} without (left) and with (right) an AC drive added to the plunger gates. A cell of the honeycomb has been overlaid for clarity. \textbf{(e)} Plot showing a direct comparison between the DC and AC current behaviour. The current polarity and quantization accuracy is determined by the trajectory followed by the pump loop around the TPs.} 
\label{Fig:FigS1}
\end{figure*}

\subsection{Quantum dot characteristics}

\label{Location of quantum dots}

\begin{figure*}[!h]
\includegraphics{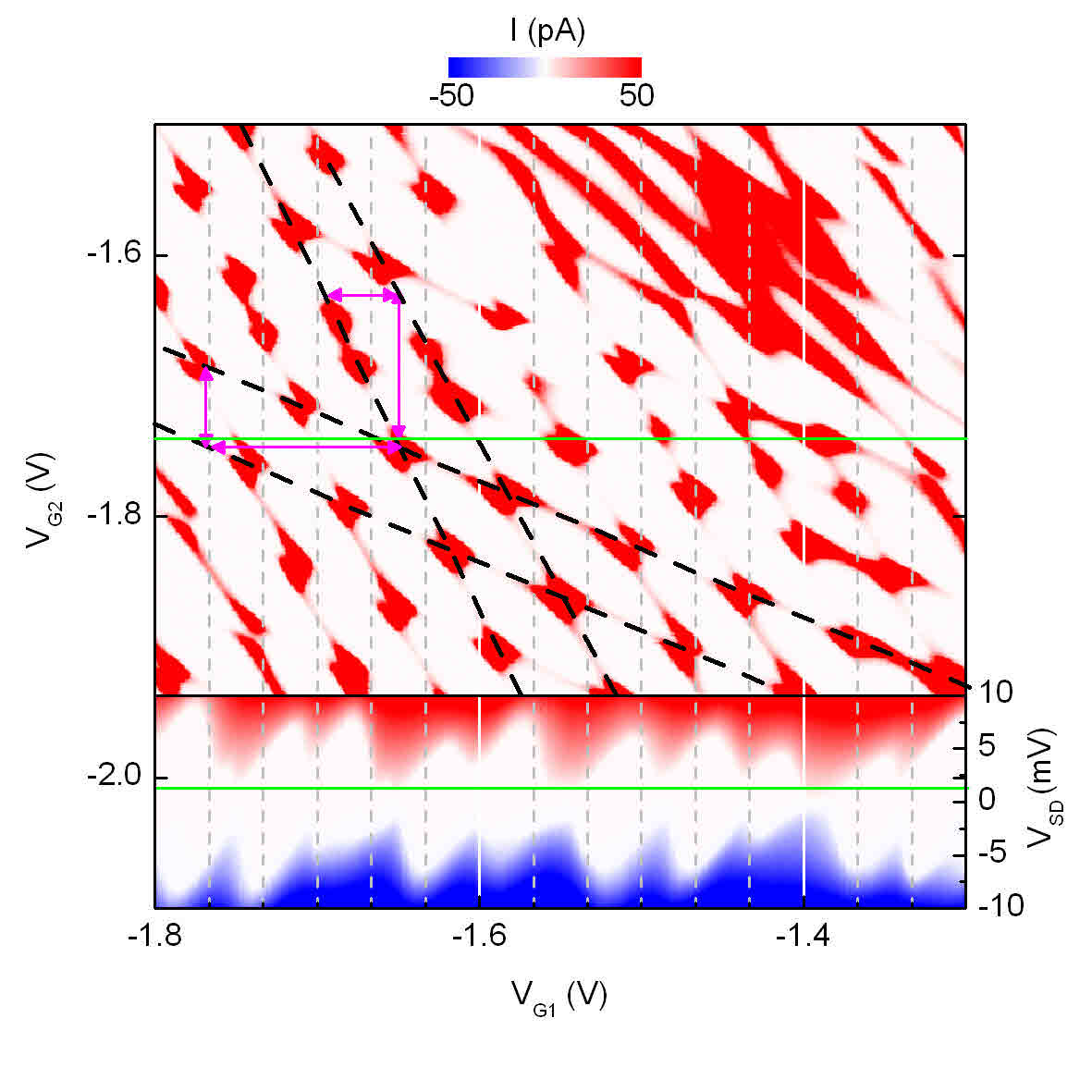}
\caption{\textbf{Stability diagram at finite bias (upper panel) and Coulomb diamonds (lower panel).} Upper green line shows the gate voltage where the lower panel was measured and the lower green line shows the source-drain bias voltage at which the upper panel was measured. Pink lines indicate the separation in gate-voltage between consecutive charge states. }    
\label{Fig:FigS4}
\end{figure*}

In order to characterize the DQD structure we measure the well developed honeycomb structure and bias triangles at a finite bias of $2$ mV, and the source-drain gap as a function of $V_{G1}$ in Fig. S\ref{Fig:FigS4}. The maximum source-drain bias gap is $\Delta V_{SD}\approx$6 mV. Assuming that the DQD can be modelled as a three junction structure, we expect the total capacitance of each dot $C_{\Sigma}$ to be related to the source-drain gap by $\Delta V_{SD}$=$3e/2C_{\Sigma}$. From this we obtain $C_{\Sigma}\approx$40 aF, which is typical of graphene QDs in the 100 nm size range \cite{Molitor2011}. The average gate-voltage separation between adjacent charge states is $\Delta V_{G1}$=48 mV ($C_{G1}$=3.3 aF) and $V_{G2}$=59 mV ($C_{G2}$=2.7 aF), confirming that the dominant contribution to the capacitance is from the source-drain contacts. To test whether these measured capacitances are consistent with dots in the lithographically designed positions, we performed electrostatic simulations using COMSOL multiphysics to calculate the capacitances and found good agreement with our observations.

\subsection{Role of disorder}
\label{Role of disorder}
Strong edge and potential disorder in lithographically defined graphene nanostructures leads to the formation of multiple quantum dots in the constrictions acting as tunnel barriers between the dots \cite{Han2007,Molitor2009,Todd2009,Stampfer2009,Liu2009,Gallagher2010,Terres2011,Connolly2011a}.
Multiple tunnel junctions (MTJs) between localized states were also observed in conventional metal-island charge pumps, and has in the past served two active roles in single electron turnstiles and pumps. Firstly, since turnstile devices only require a Coulomb gap either side of the QD, disordered MTJs can be used despite the fact that their precise form and distribution is not known \cite{Yokoi2010,Ikeda2006,Jalil1998,Siegle2010}. The second motivation for incorporating MTJs in pumps stems from the fact that cotunneling is suppressed in proportion to the number of barriers between the source and drain. For metrological applications this is particularly important as unwanted leakage current due to cotunneling destroys quantization accuracy \cite{Jensen1992,Martinis1994,Keller1996}. However, because such pumps are operated by clocking an electron sequentially through adjacent dots, individual gates must be designed to couple strongly to a specific island in the MTJ, and the randomly located junctions induced by disorder are therefore unsuitable. The high performance of the R-pump, which achieves the same error rate as MTJ pumps by fabricating on-chip resistors inline with the pump to dissipate energy and suppress cotunneling \cite{Lotkhov2001}, implies that the disorder commonly observed in graphene would improve on the basic pump as it creates a more resistive environment during pumping.  

\subsection{Oscillations in sign of pumped current with frequency}

\label{Oscillations in sign of pumped current with frequency}
\begin{figure*}
\includegraphics{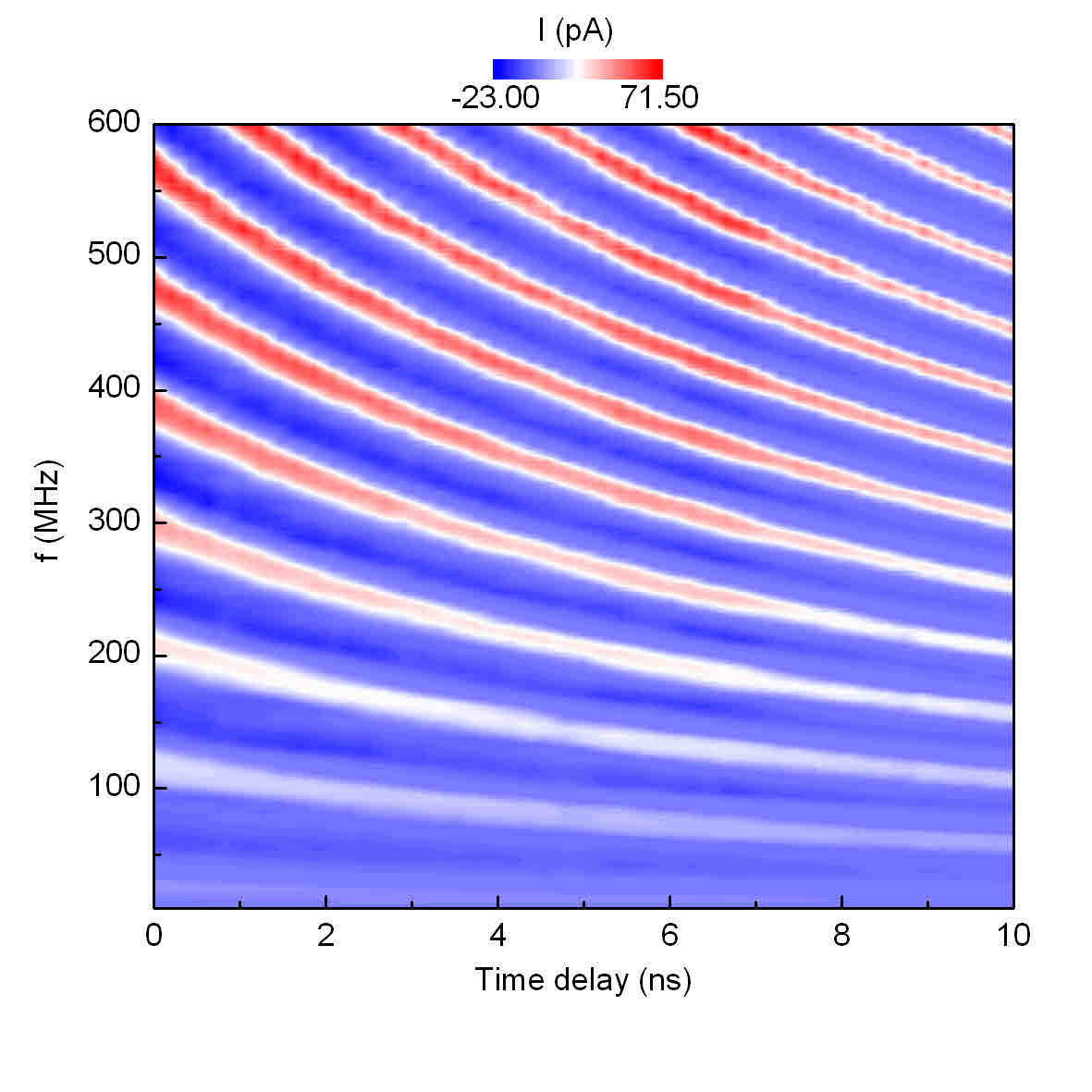}
\caption{\textbf{Pumped current as a function of frequency and the added time delay between the oscillatory components of $V_{G1}$ and $V_{G2}$.}}   
\label{Fig:FreqVsTimeDelay}
\end{figure*}

To encircle a triple point in configuration space we use a single RF generator to modulate the voltages on each gate [Fig. S\ref{Fig:FigS1}(a)]. Owing to the different cryostat wiring and circuitry used to control each gate, at a given frequency there is a built-in phase difference which leads to an arbitrary shape of the pump loop. The path length difference to each gate and non-identical frequency response of components, such as the bias tee, introduces a frequency dependence to this phase difference. Consequently, as the frequency is swept in Fig. \ref{Fig:Fig1}(c), the phase periodically walks-through $2\pi$ every time the frequency increases by 50 MHz, leading to a corresponding oscillation in the direction of the pump loop and the sign of the pumped current. In order to compensate for this and to tune the pump loop shape for optimal pumping, we introduce an additional time delay to the RF component of $V_{G2}$ using a programmable delay line [Fig. S\ref{Fig:FigS1}(a)]. Fig. S\ref{Fig:FreqVsTimeDelay} shows the pumped current as a function of frequency and time delay. The observed decrease in the period for current reversal with increasing time delay is entirely consistent with our description. 

\subsection{Dirac point location}

\begin{figure*}[!h]
\includegraphics{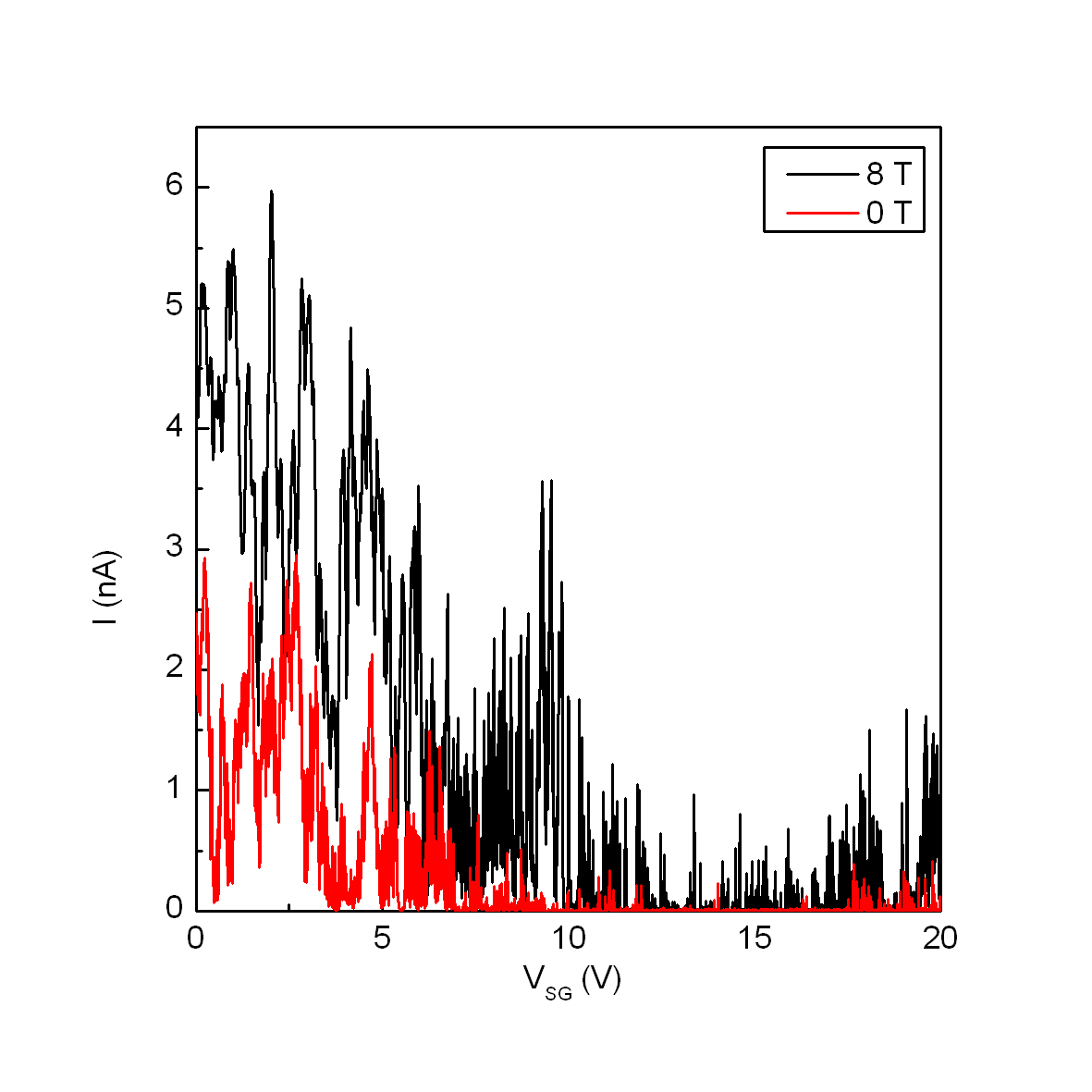}
\caption{\textbf{Current as a function of back-gate voltage showing the suppression of conductance around the Dirac point with and without an out-of-plane magnetic field ($T=$1.2 K).}}    
\label{Fig:BackgateSweeps}
\end{figure*}

\begin{figure*}[!h]
\includegraphics{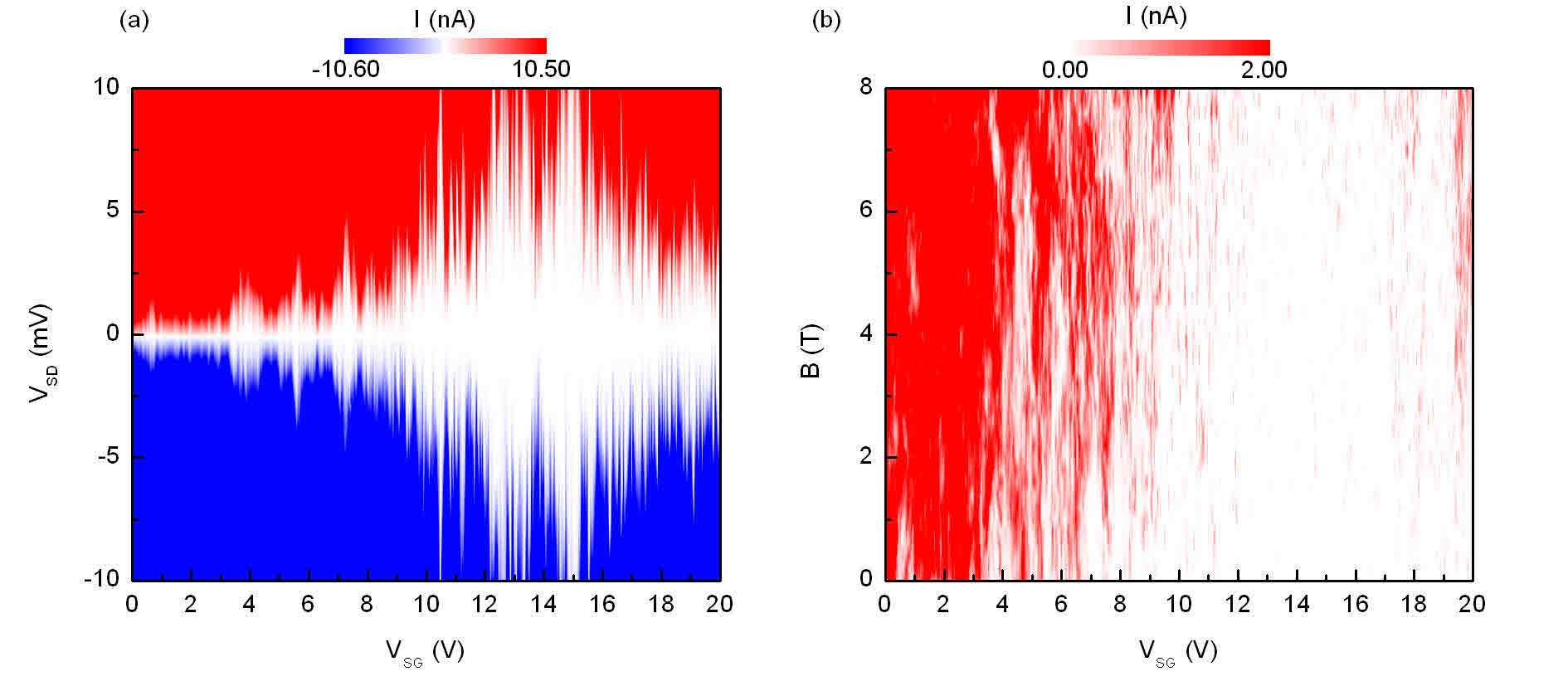}
\caption{\textbf{Source-drain bias and magnetic field dependence of transport gap ($T=$1.2 K).}}    
\label{Fig:SourceDrainMag}
\end{figure*}

In the absence of a native bandgap, charge localization in graphene relies on lateral confinement to create tunnel barriers and a combination of potential and edge disorder to produce longitudinal confinement. Fig. S\ref{Fig:BackgateSweeps} shows the source-drain current measured at $T$=1.2 K as a function of voltage applied to the side gate. The overall form of $I(V_{SG})$ has the `V' shape expected for graphene, but with a region of suppressed conductance around the Dirac point, which is characteristic of graphene nanostructures at low temperature. The fact that the Dirac point is close to $V_{SG}\approx$ 14 V suggests that there is some net charged-impurity hole doping. The transport gap $\Delta V_{SG}\approx$10 V reflects the presence of large disorder potential fluctuations. Indeed, within the transport gap there are many irregularly spaced peaks in conductance due to the alignment of energy levels in disorder-induced localized states. The energy scale of the disorder in our device can be estimated from the corresponding range of Fermi energy (in monolayer graphene) $\Delta E_F=\hbar \nu_F \sqrt{2\alpha/|e|)}$, where $\alpha$ is the side-gate capacitance per unit area, and $\nu_F$ = 10$^6$ ms$^{-1}$ is the Fermi velocity \cite{Novoselov2004}. $\Delta E_F \approx$ 60 meV is typical for exfoliated GNRs on SiO$_2$ substrates. Additional measurements confirming the location of the Dirac point are presented in Fig. S\ref{Fig:SourceDrainMag}, which shows the side-gate voltage dependence of the Coulomb gap and the magnetic field dependence of the transport gap \cite{Chiu2012}. The peak of the characteristic envelope of the charging energy and the field dependence of the transport gap symmetrically closes around $V_{SG} \approx$14 V, indicating this is the Dirac point \cite{Stampfer2009, Bai2010}. We found it was possible to pump over a range of $\approx2$V either side of the Dirac point.

\subsection{Device fabrication}
Our device is fabricated from a graphene flake obtained by mechanical exfoliation of natural graphite onto a Si substrate capped with a 300 nm thick SiO$_2$ layer. We use an undoped Si substrate in order to minimise stray capacitance, which is crucial for operating the pump at high frequency. The DQD structure, comprising two $\approx$200 nm diameter QDs separated by short $\approx$100 nm constrictions, is etched using e-beam lithography and an oxygen plasma. The absence of a back-gate means the electrostatic potential along the DQD is entirely controlled by the voltage applied to the graphene side gates: the voltages $V_{G1}$ and $V_{G2}$ on the two plunger gates are designed to change the occupancy of the QDs, while the three side gates on the opposite side tune the overall Fermi level. To operate the device as a charge pump, the output of a single RF generator is added to the DC bias of each plunger gate. 

\clearpage 


\end{document}